\documentclass{article}
\usepackage{spconf,amsmath,epsfig}

\title{OFDM Channel Estimation based on Adaptive Thresholding for Sparse Signal Detection}
\name{Mahdi Soltanolkotabi, Arash Amini and Farokh Marvasti
}
\address{Advanced Communication Research Institute (ACRI)\\ EE Department, Sharif University of Technology\\ \{msoltan , arashsil\}@ee.sharif.edu and marvasti@sharif.edu}
\begin{document}
\ninept
\maketitle
\begin{abstract}
Wireless OFDM channels can be approximated by a time varying filter with sparse time domain taps. Recent achievements in sparse signal processing such as compressed sensing have facilitated the use of sparsity in estimation, which improves the performance significantly. The problem of these sparse-based methods is the need for a stable transformation matrix which is not fulfilled in the current transmission setups. To assist the analog filtering at the receiver, the transmitter leaves some of the subcarriers at both edges of the bandwidth unused which results in an ill-conditioned DFT submatrix. To overcome this difficulty we propose Adaptive Thresholding for Sparse Signal Detection (ATSSD). Simulation results confirm that the proposed method works well in time-invariant and specially time-varying channels where other methods may not work as well.
\end{abstract}
\begin{keywords}
Sparse channel, Adaptive thresholding, OFDM channel estimation, MMSE
\end{keywords}
\section{Introduction}
\label{sec:intro}

Recent standards such as \cite{DVB-Tstandard, T-DMBstandard, ISDB-Tstandard} introduce OFDM communication as one of the best options for wireless transmission of multimedia signals including video. OFDM transmission, although resistant against multipath fading, requires accurate estimation of the Channel Frquency Response (CFR) at the receiver for appropriate decoding of the data. 

For mobile reception, the transmitter reserves some of the subcarriers in each OFDM symbol (which may vary in different symbols) for a predefined pattern of data, referred to as $Pilot$ tones, which are a priory known at the receiver. Since in OFDM transmission the data are directly sent in the frequency domain, these pilot tones provide the receiver with irregular noisy samples of the CFR in the time interval of that symbol. The main task of the channel estimator block of the receiver is to estimate CFR at the non-pilot subcarriers using the obtained samples at pilot tones. Conventional methods involve some sort of interpolation between the obtained samples, ranging from the simple Linear Interpolation (LI) to more complicated approaches like splines. Under limited time spread and Doppler frequency, the time varying OFDM channel can be viewed as a lowpass 2D signal \cite{2DSamplingOFDM}. In this sense the problem of channel estimation becomes equivalent to the reconstruction of a 2-D lowpass signal from its irregular samples \cite{Marvasti}.

Recent results in sparse signal processing such as Compressed Sensing \cite{CandesCS}, paved the way to exploit the inherent time domain sparsity of the channel in its estimation \cite{RedPilot,SparseChan2005,UnderWater}. Using $l_1$ minimization algorithms (Basis Pursuit) introduced in \cite{CandesL1}, reduction of the number of required pilots for channel estimation is proposed in \cite{RedPilot} while in \cite{UnderWater} a sparsity-based estimation method employing the Matching Pursuit is devised. 

A main drawback of these methods is that they do not consider zero padding in their scenario. In current OFDM standards the bandwidth is not fully occupied; a number of the subcarriers at both edges of the bandwidth are set to zero (hence the name zero padding) to increase the allowable transition band of the analog bandpass filter at the receiver. Zero padding results in an unstable frequency to time transformation (ill-conditioned transformation matrix). In addition, due to the lack of pilot in zero padded parts, common time domain techniques are impractical. In this paper we propose a sparse channel estimation method called Adaptive Thresholding for Detection of Sparse Signals (ATSSD) that works adequately even for zero-padded OFDM systems. Simulation results using the DVB-H standard confirm that ATSSD estimates almost the exact CFR in time-invariant channels. Furthermore, unlike the previous methods, the performance under varying channels degrades only slightly as the Doppler frequency increases.

\section{Sparse OFDM Channel Estimation }\label{sec:ProbStatement}

To show how the aforementioned challenges in OFDM channel estimation can be overcome using the inherent channel properties, we will briefly explain the problem in two subsections. In \ref{subsec:OFDMSys} OFDM system components related to channel estimation are reviewed. In \ref{subsec:Restatement}, in addition to restating the OFDM channel estimation as a sparse signal processing problem, we will explain the difficulty of partial use of bandwidth (zero padding) in estimating sparse channels.

\subsection{Review of OFDM System Components}\label{subsec:OFDMSys}

At the OFDM transmitter, coded binary data are interleaved, grouped and mapped to constellation points based on a specific modulation scheme. The resulting complex symbols which will form the frequency domain of the transmitting signal are rearranged into blocks and for each block a number of pilots are inserted among the data at predefined locations. To facilitate the RF analog filtering at the receiver, these blocks are zero padded to form the final blocks (OFDM symbols, denoted by $\mathbf X^{(n)}$ where $n$ is the block number); i.e., the bandwidth is not fully occupied. The resulting symbols are converted into the time domain ($\mathbf x^{(n)}$) by the use of the IFFT algorithm, where each symbol is extended by a Cyclic Prefix (CP, a copy of the last part of each OFDM symbol) and these final blocks are serially transmitted. The cyclic prefix is inserted to prevent possible Inter-Symbol Interference (ISI) in existence of multipath channels.

At the receiver, the cyclic prefix is removed and the received signals, $\mathbf{y}(n)$, are sent to an FFT block. When the duration of the channel impulse response is shorter than the cyclic prefix length, no ISI is expected between OFDM symbols. Assuming also no synchronization error, the equation relating the $n^{th}$ transmitted and received symbols, respectively $\mathbf{X}(n)$ and $\mathbf{Y}(n)$, is:
\begin{eqnarray}
\mathbf{Y}(n)=\mathbf{X}(n)\odot\mathbf{H}(n) + \mathbf{W}(n)
\end{eqnarray}
where $\odot$ represents element by element multiplication of two vectors, $\mathbf{H}(n)$ is a vector containing the samples of the Channel Frequency Response (CFR) corresponding to the $n^{th}$ OFDM symbol and $\mathbf{W}$ denotes the sampled vector of the AWGN noise in the frequency domain. Since the receiver is priorly aware of pilot positions and values, it can obtain a noisy estimate of the channel frequency spectrum at these subcarriers. In this sense, the OFDM channel is similar to a 2-D lattice in the discrete time-frequency plane from which certain points are known due to pilots; the goal of channel estimation is to estimate the rest via interpolation. The interpolated channel estimate is then used in the equalization block to obtain an estimate of the transmitted constellation points ($\mathbf X(n)$). After equalization, the approximated OFDM symbol is demodulated, deinterleaved, and decoded to produce the binary output data.

\subsection{Restating OFDM Channel Estimation as a Sparse Problem}\label{subsec:Restatement}

Considering the sparse distribution of the scattering objects, the OFDM channel becomes sparse in the time domain. Thus, by exploiting this sparsity, a better estimate could be obtained using the time domain.  The resulting estimate is then transformed into the frequency domain by use of the FFT algorithm. In this sense, the problem of channel estimation becomes equivalent to finding the sparse Channel Impulse Response (CIR) ($\mathbf{h}$) from the equation:
\begin{eqnarray} \label{equ:Asli}
\widetilde{\mathbf{H}}_{p}=\mathbf{F}_{p}\cdot\mathbf{h}+\mathbf{W}_{p}
\end{eqnarray}  
where $\widetilde{\mathbf{H}}_{p}$ is the vector of observed channel coefficients at pilot subcarriers, $\mathbf{F}_{p}$ is the sub-matrix of the DFT matrix obtained by keeping the rows of the FFT matrix that correspond to pilot positions and $\mathbf{W}_{p}$ is the frequency-domain noise vector at pilot positions. As stated earlier, in the case of no ISI, the length of the channel cannot exceed the length of the cyclic prefix ($N_{CP}$). Thus, (\ref{equ:Asli}) can be further simplified in the zero ISI case; only the first $N_{CP}$ elements of $\mathbf{h}$ ($\mathbf{h}_{CP}$) can have non-zero values. 
\begin{eqnarray}\label{Fari}
\widetilde{\mathbf{H}}_{p}=\mathbf{F}_{p,CP}\cdot\mathbf{h}_{CP}+\mathbf{W}_{p}
\end{eqnarray}  
where $\mathbf{F}_{p,CP}$ is the sub-matrix of the DFT matrix obtained by keeping only the first $N_{CP}$ columns of $\mathbf{F}_{p}$.
Recently the idea of using time domain sparsity in OFDM channel estimation was proposed in \cite{RedPilot} for decreasing the number of pilots. The authors proposed compressive sensing algorithms to find the sparsest time domain channel. They proved that in case of uniform pilot insertion, the matrix $\mathbf{F}_{p,CP}$ in (\ref{Fari}) satisfies the uniform uncertainty theorem described in \cite{CandesCS}. As a result, linear-programming-based algorithms used in compressed sensing, similar to the ones introduced in \cite{CandesL1} can be applied to OFDM channel estimation. However, the authors of \cite{RedPilot} did not consider zero-padding at the endpoints of the bandwidth in their scenario, which is an essential part of current OFDM standards. This assumption, causes the matrix $\mathbf{F}_{p,CP}$ to contradict the Restricted Isometric Property (RIP) defined in \cite{CandesCS} and thus the use of Compressive Sensing (CS) algorithms as described in \cite{RedPilot}, unpractical. In the presence of zero-padding, the matrix $\mathbf{F}_{p,CP}$ becomes ill-conditioned (a small variation in $\widetilde{\mathbf{H}}_p$ leads to a great change in $\mathbf{h}_{CP}$). Due to zero-padding we do not have any pilots in the zero padded parts, further complicating the use of time-domain techniques.

\section{Proposed Method}\label{ourChanEst}

As previously stated, the OFDM Channel Impulse Response (CIR) is sparse.  In this section we will propose a new scheme called Adaptive Thresholding for Sparse Signal Detection (ATSSD) that can exploit this inherent sparsity even with ill-conditioned matrices, $\mathbf{F}_{p,CP}$.

In this method, the spectrum of the channel is initially estimated using linear interpolation between pilot subcarriers. The zero-padded parts of the bandwidth which are possibly measured as nonzero due to the noise, are initially estimated as zero while at the middle subcarriers which are equipped with comb-type pilots, the mentioned linear interpolation using the noisy samples of the channel at pilot locations is employed. This initial estimate is improved in a series of iterations that finds the sparsest time domain response. 

The initially estimated spectrum is passed through the IFFT block to obtain a crude version of the time domain impulse response. This initial estimate is fed to the ATSSD block which consists of several iterations. In each iteration this method tries to find the location of the taps via a thresholding scheme on the estimated channel from the previous iteration (or in the case of the first iteration, the crude version discussed above). The thresholding scheme is further described in sec. \ref{subsec:Detection}. 
After finding the location of the taps in each iteration (locations whose previously estimated amplitudes stay above the threshold), their respective amplitudes are again found using the Minimum Mean Square Error (MMSE) criterion discussed in sec. \ref{subsec:MMSE}.  In each iteration, due to thresholding, some of the fake taps which are noise samples whose amplitudes were above the threshold in the previous iteration, are discarded. Thus, the new iteration starts with a lower number of fake taps. Moreover, because of the MMSE estimator, the valid taps approach their actual values in each new iteration. In the last iteration, the actual taps are detected and the MMSE estimator gives their respective values.  
The main steps of the proposed algorithm are:
\begin{enumerate}
\item Using linear interpolation, set the crude time-domain channel as described above for the initial estimate.
\item Discard the taps with amplitudes bellow the threshold.   
\item Estimate the value of the remaining taps using the MMSE criterion.
\item Stop if the estimated channel is the same as the previous iteration or when a maximum number of iterations is reached, else go to step 2.  
\end{enumerate}

\subsection{Channel Tap Detection via Thresholding}\label{subsec:Detection}

As mentioned earlier, in each iteration, taps bellow a certain threshold are discarded. The threshold in the $i^{th}$ iteration is set as:
\begin{eqnarray}\label{eq:ThreshMethod}
Threshold=\beta e^{\alpha i} 
\end{eqnarray} 
where $i$ represents the iteration number; i.e., the threshold exponentially increases as the iterations proceed. $\alpha$ and $\beta$ are constants that depend on the number of taps and initial powers of noise and channel taps; since these parameters are hardly known, a rough approximation is used. In the first iteration, the threshold is a small number and with each iteration it is gradually increased. Intuitively, this gradual increase of the threshold, results in a gradual reduction of fake taps (taps that are created due to noise). This is further explained in section \ref{MathAnalaysis} from a mathematical point of view.

\subsection{Estimation of Channel Tap Values via MMSE}\label{subsec:MMSE}

After finding time-domain channel tap positions, in each iteration we need to find an estimate of their value. This is equivalent to solving a linear equation in the presence of noise. That is, we wish to obtain the value of the CIR vector $\mathbf{h}$ at tap positions ($T$), from the equation
\begin{eqnarray}
\widetilde{\mathbf{H}}_{p}=\mathbf{F}_{p,T}\mathbf{h}_T+\mathbf{W}_{p}
\end{eqnarray}
where the vector $\widetilde{\mathbf{H}}_{p}$ is the measured CFR vector at pilot positions, $\mathbf{F}_{p,T}$ is the sub-matrix of the DFT matrix ($\mathbf{F}$) obtained by selecting the rows and columns that pertain to the pilot positions ($p$) and likely channel taps ($T$), respectively, and $\mathbf{W}_{p}$ is the noise vector at pilot positions. The MMSE estimator tries to find the estimate vector $\hat{\mathbf{h}}_T$ that minimizes $E\{\|\mathbf{h}_T-\hat{\mathbf{h}}_T\|_2\}$, hence the name Minimum Mean Squared Error. This estimate vector is given by:
\begin{eqnarray}
\hat{\mathbf{h}}_T = \mathbf{R}_{\mathbf{h}_{T}} \mathbf{F}_{p,T}^H \big(\mathbf{F}_{p,T}\mathbf{R}_{\mathbf{h}_{T}}\mathbf{F}_{p,T}^H + \mathbf{R}_\mathbf{W}\big)^{-1}\widetilde{\mathbf{H}}_{p}
\end{eqnarray}   
where $.^H$ denotes the Hermitian operation and $\mathbf{R}_{\mathbf{h}_{T}}$ and $\mathbf{R}_\mathbf{W}$ are the auto-covariance matrices of $\mathbf{h}_T$ and $\mathbf{W}_{p}$ respectively. When the noise vector $\mathbf{W}$ is a random complex white Gaussian process, $\mathbf{R}_\mathbf{W}$ is equal to $2\sigma_\mathbf{W}^2\mathbf{I}$, where variance of both real and imaginary parts are assumed to be $\sigma_\mathbf{W}^2$. Also $\mathbf{R}_{\mathbf{h}_{T}}$ can be approximated with $P_\mathbf{h}\mathbf{I}$ where $P_\mathbf{h}$ is the average power of the channel. Thus, the estimate vector can be written as:
\begin{eqnarray}
\hat{\mathbf{h}}_T \approx \mathbf{F}_{p,T}^H \big(\mathbf{F}_{p,T}\mathbf{F}_{p,T}^H +\frac{2\sigma_\mathbf{w}^2}{P_\mathbf{h} } \mathbf{I}\big)^{-1}\widetilde{\mathbf{H}}_{p}
\end{eqnarray}
where the channel power $P_h$ can easily be estimated by dividing the power of the to-be-equalized data by the constellation power (power of the equalized data is the same as the average constellation power).

\section{Mathematical Analysis}\label{MathAnalaysis}

In this section we present a mathematical intuition for the proposed method. For simplicity of the analysis, we assume that both channel taps and the additive noise in the time domain at the $i^{th}$ iteration (including the initial iteration formed by linear interpolation) are zero-mean normal complex random variables with equal real and imaginary variances $\sigma_{tap}^2$ and $\sigma_{n,i}^2$, respectively. Therefore, their amplitudes which are of main concern in ATSSD have Rayleigh probability distribution function. In addition, we assume that the probability of a time sample being a channel tap is $p_{tap}$; i.e., if $N_{CP}$ represents the length of the cyclic prefix, we expect to have a channel with $p_{tap}\cdot N_{CP}$ taps. Now, if we set a threshold ($\eta$) to distinguish the noise and valid samples, we probably miss some of the original taps while detecting a number of fake locations. The probability of False Alarm (FA); i.e., detection of fake taps,using the Rayleigh distribution is given by:
\begin{eqnarray}
p_{fa}=\int_{\eta}^{\infty}\frac{x}{\sigma_{n,i}^{2}}e^{-\frac{x^2}{2\sigma_{n,i}^2}} \cdot dx=e^{-\frac{\eta^2}{2\sigma_{n,i}^2}}
\end{eqnarray}
Moreover, the average power of these fake taps ($\Sigma_{tap|n}$) is:
\begin{eqnarray}
\Sigma_{tap|n} &=& E\big\{|w|^2~~\big|~~|w|>\eta\big\} \nonumber \\
&=&\frac{1}{p_{fa}} \int_{\eta}^{\infty}\frac{x^3}{\sigma_{n,i}^{2}}e^{-\frac{x^2}{2\sigma_{n,i}^2}} \cdot dx\nonumber\\
&=&\frac{1}{p_{fa}}2\sigma_{n,i}^{2}\big(1+\frac{\eta^2}{2\sigma_{n,i}^{2}}\big) e^{-\frac{\eta^2}{2\sigma_{n,i}^{2}}}\nonumber\\
&=& 2\sigma_{n,i}^{2}\big(1+\frac{\eta^2}{2\sigma_{n,i}^{2}}\big)
\end{eqnarray}
Assuming $\eta$ as the threshold for the $i^{th}$ iteration, we expect the overall noise power (real and imaginary) of the next iteration to be equal to $\Sigma_{tap|n}$. Although this is only a rough statistical approximation, simulation results confirm that after MMSE estimation, the noise in the remaining taps (locations that stayed above the threshold in the previous iteration) is almost normal with zero mean and variance $0.5\Sigma_{tap|n}$:
\begin{eqnarray}{\label{eq:SigmaNext}}
\sigma_{n,i+1}^{2}\approx \frac{\Sigma_{tap|n}}{2}\bigg|_{\eta}= \sigma_{n,i}^{2}\big(1+\frac{\eta^2}{2\sigma_{n,i}^{2}}\big)
\end{eqnarray}
For the choice of $\eta$, in \cite{SinaJournal} it is shown that the threshold which maximizes SNR of the signal after thresholding (considering both missed and fake taps) is found as:
\begin{eqnarray}\label{eq:OptThresh}
\eta_{opt,i}=\sigma_{n,i}^2\sqrt{2\frac{\sigma_{n,i}^2+\sigma_{tap}^2}{\sigma_{tap}^2} \ln\bigg(\frac{1-p_{tap}}{p_{tap}} \frac{\sigma_{n,i}^2+\sigma_{tap}^2}{\sigma_{n,i}^2}\bigg)}
\end{eqnarray}
Using (\ref{eq:OptThresh}) in (\ref{eq:SigmaNext}) we get:
\begin{eqnarray}
\frac{\sigma_{n,i+1}^{2}}{\sigma_{n,i}^{2}}=1+ \frac{\sigma_{n,i}^2+\sigma_{tap}^2}{\sigma_{tap}^2} \ln\bigg(\frac{1-p_{tap}}{p_{tap}} \frac{\sigma_{n,i}^2+\sigma_{tap}^2}{\sigma_{n,i}^2}\bigg)
\end{eqnarray}
which shows the increase in the noise variance; the noise samples which stay above the threshold should have higher amplitudes and therefore, higher variance. For reasonable SNR values of the channel, $\frac{\sigma_{tap}^2}{\sigma_{n,i}^2}$ is large enough to have the following approximations:
\begin{eqnarray}\label{eq:ApproxThresh1}
\left\{\begin{array}{l}
\eta_{opt,i}\approx \sigma_{n,i}^2\sqrt{2 \ln\bigg(\frac{1-p_{tap}}{p_{tap}} \frac{\sigma_{tap}^2}{\sigma_{n,i}^2}\bigg)}\\
\sigma_{n,i+1}^{2}\approx\sigma_{n,i}^{2}\bigg(\underbrace{1+\ln\big(\frac{1-p_{tap}}{p_{tap}} \frac{\sigma_{tap}^2}{\sigma_{n,0}^2}\big)}_{k}-\ln(\frac{\sigma_{n,i}^2}{\sigma_{n,0}^2})\bigg)
\end{array}\right.
\end{eqnarray}

As can be seen in (\ref{eq:ApproxThresh1}), $\sigma_{n,i+1}^{2}<k\sigma_{n,i}^{2}$; however, we assume the upper bound for the noise and thus:
\begin{eqnarray}\label{eq:MathResult}
\sigma_{n,i+1}^{2}&\approx& k\sigma_{n,i}^{2} \Rightarrow \sigma_{n,i}^{2} \approx \sigma_{n,0}^{2} k^i \nonumber\\
\Rightarrow \eta_{opt,i}&\approx&\sigma_{n,0}^{2} e^{i\ln(k)}\sqrt{2\ln\big(\frac{1-p_{tap}}{p_{tap}}\sigma_{tap}^2\big)-4i\ln(k)}\nonumber\\
&\approx& \beta e^{i\ln(k)}
\end{eqnarray}
where $\beta$ is a constant. Similar to (\ref{eq:ThreshMethod}), (\ref{eq:MathResult}) shows the same exponential increase in the threshold with respect to the iteration number.

\section{Simulation Results}\label{SimResults}

\begin{figure}[htb]

\begin{minipage}{1\linewidth}
  \centering
  \centerline{\epsfig{figure=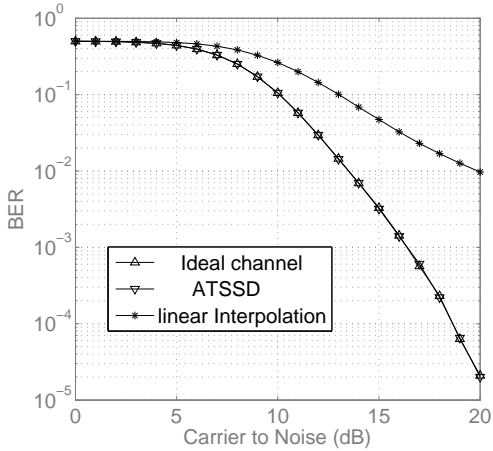,width=7cm}}
\end{minipage}
\hfill
\caption{Performance of the ideal, linear-interpolated and the proposed channel estimates under time-invariant Brazil D channel.}
\label{fig:BERAV}
\end{figure}

\begin{figure}[htb]

\begin{minipage}{1\linewidth}
  \centering
  \centerline{\epsfig{figure=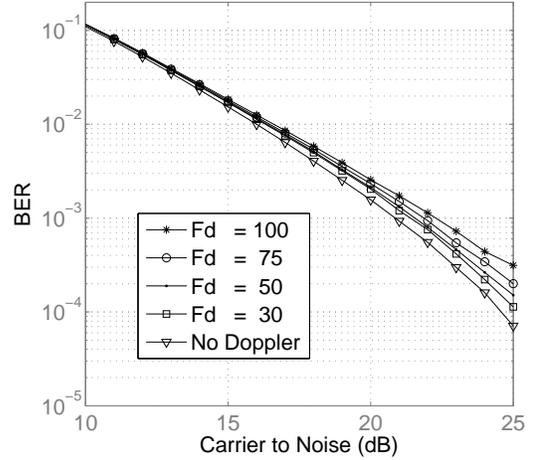,width=7.4cm}}
\end{minipage}
\hfill
\caption{Performance of the proposed method at different Doppler frequencies for Brazil D channel.}
\label{fig:BERDoppler}
\end{figure}

To verify the efficacy of the proposed channel estimation, we performed computer simulations based on the DVB-H standard using MATLAB. The main simulation parameters and options are shown in Table \ref{tab:DVB} and \ref{tab:BrazilProf}. 
To show the robustness of ATSSD in estimating the CFR for time-invariant channels, we compared the BER after Viterbi decoding of the proposed method to that of the hypothetical ideal channel in Fig. \ref{fig:BERAV}; by ideal channel we mean that for data equalization, we used the exact channel instead of its estimate. As can be seen in Fig. \ref{fig:BERAV}, the BER performance of the proposed method almost coincides with that of the ideal channel. Thus, in this sense, the ATSSD estimation is perfect. Moreover, the BER performance of Linear Interpolation (a conventional channel estimation method) is also shown for comparison. The proposed method is also effective in time-varying channels. This fact is verified in Fig. \ref{fig:BERDoppler} where the BER performance (after Viterbi decoding) of ATSSD is shown for different Doppler frequencies. As obvious in Fig. \ref{fig:BERDoppler}, ATSSD shows little performance degradation with increase of the Doppler frequency.

\begin{table}[tb]
\begin{center}
\caption{Simulation Parameters}
\begin{tabular}[tb]{c c c c}
\hline \hline
Parameter & & & Specifications \\
\hline\hline
 DVB-H mode & & & 2K \\
\hline
 Number of carriers & & & $1705$ \\ 
\hline
 Number of pilot carriers & & & $146$ \\
\hline
 OFDM symbol duration & & & $224 \mu s$ \\
\hline
 Gaurd Interval & & & $1/8~(256)$ \\ 
\hline
 Signal Constellation & & & QAM-16\\
\hline
Convolutional Encoder & & & $171_{OCT}$ and $133_{OCT}$  \\
\hline
 Coding rate & & & 1/2  \\
\hline
$[iter_{max}~,~\alpha~,~\beta]$ & & & $[5~,~0.8~,~0.008]$  \\
\hline \hline
\label{tab:DVB}
\end{tabular}
\end{center}
\end{table}


\begin{table}[tb]
\begin{center}
\caption{Multi-Path Profile (Brazil Channel D)}
\begin{tabular}[tb]{c c c c c c c c}
\hline \hline
Delay ($\mu s$) & $0.0$ & $0.48$ & $2.07$ & $2.90$ & $5.71$ & $5.78$ \\
\hline
Amp. (dB) & $-0.1$ & $-3.9$ & $-2.6$ & $-1.3$ & $0.0$ & $-2.8$ \\
\hline
\label{tab:BrazilProf}
\end{tabular}
\end{center}
\end{table}

\section{Conclusion}

Due to the partial use of bandwidth in OFDM transmission, the reduced frequency to time transformation is no longer stable; therefore, most of the sparsity-based channel estimation methods diverge.
We have proposed a thresholding method which solves this problem while benefiting from the sparsity criterion. Simulation results show that the performance of this method after data equalization is almost the same as if
we used the exact channel even for time-varying channels.

\bibliographystyle{IEEEbib}
\bibliography{strings,refs}

hamash alakie


\bibitem{RedPilot,
G. Taubock and F. Hlawatsch, "A compressed sensing technique
for ofdm channel estimation in mobile environment: Exploiting
channel sparsity for reducing pilots," in ICASSP2008.
IEEE, 2008, pp. 2885–2888.}

\bibitem{UnderWater,
T. Kang and R. Iltis, "Matching pursuits channel estimation for
an underwater acoustic ofdm modem," in ICASSP2008. IEEE,
2008, pp. 5296–5299.}


\bibitem{CandesCS,
E. Candes, J. Romberg, and T. Tao, "Stable signal recovery
from incomplete and inaccurate measurments," Comm. Pure
Appl. Math., vol. 59, pp. 1207–1223, March 2006.}

\bibitem{2DSamplingOFDM,
F. Sanzi and J. Speidel, "An adaptive two-dimensional channel
estimator for wireless ofdm with application to mobile dvb-t,"
IEEE Trans. Broadcasting, vol. 46, no. 2, pp. 128–133, June
2000.}

\bibitem{SparseChan2005,
M. Raghavendra and K. Giridhar, "Improving channel estimation
in ofdm systems for sparse multipath channels," IEEE Sig.
Proc. Letters, vol. 12, no. 1, pp. 52–55, Jan. 2005.}

\bibitem{SinaJournal,
S. Zahedpour, S. Feizi, A. Amini, and F. Marvasti, "Impulsive
noise cancellation based on soft decision and recursion," to appear
in IEEE Trans. Instrumentation and Measurement, Dec.
2008.}

\bibitem{DVB-Tstandard,
ETSI EN 302 304, "Digital video broadcasting (dvb); transmission
system for handheld terminals (dvb-h)," Nov. 2004.}

\bibitem{ISDB-Tstandard,
ARIB STD-B29 Ver.2.1, "Transmission system for digital terrestrial
sound broadcasting," Jul. 2003.}

\bibitem{T-DMBstandard,
ETSI TS 102 428, "Digital audio broadcasting (dab); dmb
video service; user application specification," June 2005.}

\bibitem{CandesL1,
E. Candes and J. Romberg, "l1-magic: Recovery
of sparse signals via convex programming,"
http://www.acm.caltech.edu/l1magic.}

\bibitem{Marvasti,
F. Marvasti, Nonuniform Sampling: Theory and Practice,
Kluwer Academic/Plenum Publishers, 2001.}
\end{document}